\documentclass[12pt,a4paper]{article}
\usepackage{parskip}
\usepackage[OT2,T1]{fontenc}

\usepackage{amssymb}
\usepackage{amsmath}
\usepackage{stmaryrd}
\usepackage{mathrsfs}
\usepackage{mathtools}
\usepackage{bm}
\DeclareMathAlphabet{\mathscrbf}{OMS}{mdugm}{b}{n}

\usepackage{framed}
\usepackage{tikz}
\usepackage{graphicx}
\usepackage[leftcaption]{sidecap}
\usepackage{caption}
\usepackage{subcaption}
\usepackage{here}
\usepackage[margin=.95in]{geometry}

\usepackage{needspace}
\usepackage{titlesec}

\usepackage{tocloft}

\usepackage{enumerate}

\usepackage{cite}
\let\OLDthebibliography\thebibliography
\renewcommand\thebibliography[1]{
  \OLDthebibliography{#1}
  \setlength{\parskip}{0pt}
  \setlength{\itemsep}{0pt plus 0.3ex}
}

\usepackage[colorlinks,allcolors=blue]{hyperref}

\newcommand{\be}{\begin{equation}}
\newcommand{\ee}{\end{equation}}
\newcommand{\ben}{\begin{enumerate}}
\newcommand{\een}{\end{enumerate}}
\newcommand{\bi}{\begin{itemize}}
\newcommand{\ei}{\end{itemize}}
\newcommand{\bmm}{\begin{pmatrix}}
\newcommand{\emm}{\end{pmatrix}}

\newcommand{\Ad}{\text{Ad}}
\newcommand{\ad}{\text{ad}}

\newcommand{\dd}{\text{d}}

\newcommand{\der}{\partial}
\newcommand{\Diff}{\text{Diff}\,S^1}

\newcommand{\ds}{\displaystyle}

\newcommand{\eg}{{\it e.g.}\ }

\newcommand{\hDiff}{\widehat{\text{Diff}}\,S^1}

\newcommand{\ie}{{\it i.e.}\ }

\newcommand{\Vect}{\text{Vect}\,S^1}

\newcommand{\bu}{\textbf u}
\newcommand{\bx}{\textbf x}

\newcommand{\cI}{{\cal I}}

\newcommand{\cR}{{\cal R}}

\newcommand{\cV}{{\cal V}}
\newcommand{\cW}{{\cal W}}

\newcommand{\mg}{\mathfrak{g}}

\newcommand{\sfG}{\mathsf{G}}

\newcommand{\sfM}{\mathsf{M}}

\newcommand{\RR}{\mathbb{R}}
\newcommand{\ZZ}{\mathbb{Z}}

\titleformat{\section}{\normalfont\fontsize{14}{14}\bfseries}{\thesection}{1em}{}
\titleformat{\subsection}{\normalfont\fontsize{12}{12}\bfseries}{\thesubsection}{1em}{}

\begin{document}

\begin{center}
\hrule
\makebox[\textwidth][c]{\large{\bfseries{Topological Bifurcations and Reconstruction of Travelling Waves}}}
\vspace{-.6em}
\hrule
\end{center}
\vspace{.5cm}

\begin{center}
Blagoje Oblak
\end{center}
\vspace{.5cm}

\begin{center}
\begin{minipage}{.9\textwidth}\small\it
\begin{center}
{\tt{boblak@lpthe.jussieu.fr}}\\
LPTHE, Sorbonne Universit\'e and CNRS UMR 7589, F-75005 Paris, France;\\
CPHT, Ecole Polytechnique and CNRS UMR 7644, F-91128 Palaiseau, France.
\end{center}
\end{minipage}
\end{center}
\vspace{2.5cm}

\begin{center}
\begin{minipage}{.92\textwidth}
\paragraph{Abstract.} This paper is devoted to periodic travelling waves solving Lie-Poisson equations based on the Virasoro group. We show that the reconstruction of any such solution can be carried out exactly, regardless of the underlying Hamiltonian (which need not be quadratic), provided the wave belongs to the coadjoint orbit of a uniform profile. Equivalently, the corresponding `fluid particle motion' is integrable. Applying this result to the Camassa-Holm equation, we express the drift of particles in terms of parameters labelling periodic peakons and exhibit orbital bifurcations: points in parameter space where the drift velocity varies discontinuously, reflecting a sudden change in the topology of Virasoro orbits.
\end{minipage}
\end{center}
\vspace{2.5cm}

\tableofcontents
~\\

\newpage
\section{Introduction}

A standard problem in fluid mechanics is to describe the motion of particles sourced by a propagating wave \cite{Ursell}. This is embodied by the concept of  {\it reconstruction} in symplectic geometry \cite{Marsden,Ratiu,Arnold}: knowing the evolution of a system's momentum (or velocity), one has to find the corresponding motion in configuration space. The issue typically turns out to be harder than it sounds; think \eg of the intricate rotations of a rigid body \cite[sec.\ VI]{Landau}, including geometric phases \cite{Montgomery}, despite the relatively simple equations satisfied by its angular momentum. The hydrodynamical analogue of this setup is a wave equation describing the evolution of the fluid's velocity $\bu(\bx,t)$, whose reconstruction $\dot\bx=\bu(\bx(t),t)$ is equivalent to a geodesic in a diffeomorphism group \cite{Arnold,Modin} and is generally accessible only through numerical methods or asymptotic calculations \cite{Chang}.

The purpose of this paper is to point out that a broad class of solutions of fairly general nonlinear wave equations in one dimension can be reconstructed analytically, in a manner that exhibits the underlying infinite-dimensional geometry. Concretely, we consider periodic travelling waves that satisfy Lie-Poisson (or Euler-Poincar\'e\footnote{Following \cite{Marsden}, we refer to Hamiltonian equations of motion (in momentum space) as {\it Lie-Poisson equations}, while their Lagrangian analogues (in velocity space) are called {\it Euler-Poincar\'e equations}. In the literature, such systems also go under the name of Euler-Arnold or Euler-Poisson equations \cite{Marsden,Ratiu,Khesin2003,Khesin}.}) equations associated with the Virasoro group \cite{Khesin2003,Khesin}, with an arbitrary, generally non-quadratic, Hamiltonian. Provided the wave profile is {\it amenable}, \ie Virasoro-equivalent to a uniform field configuration \cite{Lazutkin,Kirillov,Witten:1987ty,Balog:1997zz}, we show that reconstruction can be carried out exactly --- without approximations --- in terms of explicit circle diffeomorphisms. This leads in particular to a simple expression for the average drift velocity of `fluid particles' solving the equation of motion $\dot x(t)=u(x(t),t)$. The quotation marks stress that the model $\dot x=u$ does {\it resemble} the motion of fluid particles in shallow water \cite{Korteweg,Camassa}, but also differs from it in key details that make the phenomenological implications of our analysis somewhat unclear.

Similar observations have recently been applied to cnoidal waves in \cite{Oblak:2019llc,Oblak:2020jek}, where simplifications specific to Korteweg-de Vries (KdV) dynamics obscured the distinction between `momentum' and `velocity'. Our goal is therefore to generalize the method to more complicated nonlinear dynamics. As an illustration, we shall consider the Camassa-Holm (CH) equation \cite{Camassa}, whose simplest periodic travelling waves are peakons or `coshoidal waves' \cite{Boyd}. Similarly to \cite{Oblak:2019llc,Oblak:2020jek}, reconstruction will relate wave parameters to Virasoro coadjoint orbits \cite{Lazutkin,Kirillov,Witten:1987ty,Balog:1997zz} and exhibit points in parameter space where orbital topology changes discontinuously. We refer to such abrupt changes as {\it topological bifurcations}, and show that they lead to discontinuities in the parameter-dependence of drift velocity.

One should note that the reconstruction of travelling waves is elementary from the standpoint of dynamical systems: as we explain in section \ref{secrec}, it is equivalent to a flow equation $\dot X(t)=U(X(t))$, where $U$ is continuous and periodic. The solution of this problem is implicitly given by $\int_{X(0)}^{X(t)}\dd X/U(X)=t$, which readily yields the value of drift velocity. It is thus no surprise that reconstruction can be carried out exactly; what is perhaps less obvious is that it can be performed explicitly in terms of circle diffeomorphisms, in a way that highlights the relation between reconstruction and Virasoro orbits. In particular, the topological transitions mentioned above are just saddle-node bifurcations of the system $\dot X=U(X)$ --- but their orbital interpretation appears to be new. Our focus on this rephrasing entails a point of view that may seem peculiar to readers acquainted with dynamical systems, but we hope this will not lead to any confusion. Prerequisites for this work include Virasoro group theory and the ensuing orbit classification \cite{Lazutkin,Kirillov,Witten:1987ty,Balog:1997zz}, as well as Lie-Poisson equations and  reconstruction. We refer \eg to \cite{Guieu} or \cite[chaps.\ 6-7]{Oblak:2016eij} for a pedagogical introduction to the former, and to \cite{Marsden,Ratiu} for the latter. See also \cite{Oblak:2019llc,Oblak:2020jek} for closely related background and motivations.

The plan is as follows. In section \ref{segen}, we expose a general reconstruction argument for travelling waves satisfying a Lie-Poisson equation, with arbitrary Hamiltonian, based on the Virasoro group. Aside from providing a formula for drift velocity, this will also allow us to establish a dictionary between one-dimensional vector flows and Virasoro orbits. Section \ref{sech} then applies this result to periodic peakons, some of which are amenable while others are not, leading to a rich bifurcation diagram. Whether the ensuing discontinuities of drift velocity can in fact be observed in experiments is a crucial question, which we leave for future work.

\section{Reconstruction of travelling waves}
\label{segen}

This section is devoted to our main result: under certain weak assumptions on the Hamiltonian, any amenable travelling wave admits an explicit Lie-Poisson reconstruction. The proof is an extension of \cite[sec.\ 4.1]{Oblak:2020jek} to arbitrary dynamics. Accordingly, we start by introducing general Lie-Poisson (or Euler-Poincar\'e) equations associated with the Virasoro group. Their reconstruction is reminiscent of an equation of motion for fluid particles, which motivates the definition of an average `drift velocity'. We then show that the reconstructed path associated with any amenable travelling wave can be written entirely in terms of basic wave data; the same is true of drift. This eventually leads to a theorem relating wave amenability to the roots of the velocity profile, and implies the existence of saddle-node bifurcations sensitive to the topology of Virasoro orbits. Applications of these ideas to the CH equation and periodic peakons \cite{Camassa,Boyd} are postponed to section \ref{sech}.

\subsection{Virasoro group and Lie-Poisson equations}
\label{sevigo}

Here we briefly recall the derivation of Lie-Poisson equations associated with Hamiltonian functionals on the dual of the Virasoro algebra. In particular, we introduce notions of `momentum' and `velocity' that will appear repeatedly below. For an introduction to the Virasoro group and its coadjoint orbits, see \eg \cite[sec.\ II.2]{Khesin}, \cite{Guieu} or \cite[chaps.\ 6-7]{Oblak:2016eij}.

\paragraph{Virasoro group and orbits.} Consider a circle $S^1$ whose points are labelled by a real coordinate $x\sim x+2\pi$. The resulting diffeomorphism group, $\Diff$, consists of smooth\footnote{In section \ref{sepek}, peakons will force us to weaken the assumption of smoothness (see footnote \ref{fosmop}).\label{fosmo}} maps $x\mapsto f(x)$ such that $f'(x)>0$ and $f(x+2\pi)=f(x)+2\pi$, where $f'\equiv\dd f/\dd x$. Each $f$ has an inverse $f^{-1}$ such that $f^{-1}(f(x))=f(f^{-1}(x))=x$. In this language, rotations read $f(x)=x+\theta$; such transformations will be important for our purposes, so from now on we write $\cR_{\theta}(x)\equiv x+\theta$. The Lie algebra of $\Diff$ consists of vector fields $u=u(x)\der_x\in\Vect$, with $u(x+2\pi)=u(x)$.

The universal central extension of $\Diff$ is the {\it Virasoro group} $\hDiff$, whose algebra has a dual space consisting of pairs $(p,c)$ with $c\in\RR$ a central charge and $p=p(x)\dd x^2\in(\Vect)^*$ a $2\pi$-periodic quadratic density, seen as a distribution that maps $u\in\Vect$ on $\int\dd x\,p(x)u(x)$.\footnote{The densities $p(x)$ of peakons will contain delta function singularities (see footnote \ref{fosmo}).\label{fosmop}} The action of $\hDiff$ on this dual space is the coadjoint representation
\be
\Ad^*_f(p,c)
\equiv
(f\cdot p,c)
=
\bigg(
\big((f^{-1})'\big)^2p\circ f^{-1}
-
\frac{c}{12}\bigg[\frac{(f^{-1})'''}{(f^{-1})'}-\frac{3}{2}\Big(\frac{(f^{-1})''}{(f^{-1})'}\Big)^2\bigg],
c\bigg),
\label{cog}
\ee
where the notation $f\cdot p$ is introduced for later convenience. The resulting coadjoint representation of the Virasoro Lie algebra reads
\be
\ad^*_{u}(p,c)
=
\Big(-up'-2u'p+\frac{c}{12}u''',0\Big).
\label{adot}
\ee
For instance, eq.\ \eqref{cog} yields $(\cR_{\theta}\cdot p)(x)=p(x-\theta)$ for rotations. More generally, the {\it coadjoint orbit} of $p$ is the subset of $(\Vect)^*$ that consists of all points of the form $f\cdot p$, where $f$ runs over $\Diff$. Integrable reconstruction will apply to $p$'s that can be mapped in this way on a uniform configuration $k(x)=\text{cst}$, so orbits will play a key role below.

We shall think of coadjoint vectors $p(x)$ as wave profiles, to be governed by Lie-Poisson evolution equations such as KdV or its cousins \cite{Khesin2003,Khesin}; from a symplectic perspective, $p$ can be seen as a momentum vector \cite{Marsden}. Conversely, from a Lagrangian perspective, Lie algebra elements $u$ are `velocities'; this interpretation is doubly valid in hydrodynamics \cite{Arnold}, where vector fields describe the actual local velocity of a fluid.

\paragraph{Hamiltonian and velocity.} Let $H[p]$ be a Hamiltonian functional of $p\in(\Vect)^*$. Throughout this work, we impose the following constraints on $H$:
\begin{enumerate}[(i)]
\item We assume that $H$ is differentiable, \ie that its functional derivative $\delta H/\delta p(x)$ exists, and that the {\it velocity map}
\be
\cW:(\Vect)^*\to\Vect:p\mapsto\cW[p]
\equiv
\cW[p](x)\der_x
\equiv
\frac{\delta H}{\delta p(x)}
\label{wmap}
\ee
is a bijection. The terminology is justified by Hamilton's equation of motion $\dot q=\der H/\der p$, and by the fact that the vector field $\cW[p]$ is typically related to the local velocity of fluid particles in shallow water. (This will motivate the definition of `drift' in section \ref{secrec}.) Letting \eqref{wmap} be a bijection ensures that any velocity $u\in\Vect$ reads $u=\cW[p]$ for some unique momentum $p\in(\Vect)^*$. In particular, there exists an inverse map $\cI:\Vect\to(\Vect)^*:u\mapsto\cI[u]$ such that $\delta H/\delta p\big|_{p=\cI[u]}=u$. For quadratic Hamiltonians, $\cI$ is in fact an inertia operator \cite{Khesin} --- hence the notation.
\item We demand that the Hamiltonian be {\it invariant under rotations} in the sense that $H[p(x)]=H[p(x-\theta)]$ for all $\theta\in\RR$. This yields, for the velocity map, the identity
\be
\cW[\cR_{\theta}\cdot p](x)
=
\cW[p](x-\theta),
\label{wrot}
\ee
which will be crucial for the reconstruction of travelling waves. Indeed, a travelling wave is one for which $p(x,t)=p(x-vt)$, and eq.\ \eqref{wrot} implies that the corresponding local velocity $u(x,t)\equiv\cW[p(t)](x)$ can be written as $u(x,t)=u(x-vt)$.
\end{enumerate}
In practice, rotational symmetry is a weak condition; it is satisfied, to our knowledge, by all Lie-Poisson equations that have appeared in the literature so far \cite{Arnold,Khesin,Modin}. For example, the CH Hamiltonian is a non-local functional $H[p]\propto\iint\dd x\,\dd y\,\sfG(x-y)p(x)p(y)$, where $\sfG$ is a certain Green's function (see section \ref{sepege}). The velocity map thus reads $\cW[p](x)=\int\dd y\,\sfG(x-y)p(y)$ and reduces to $\cW[p](x)=p(x)$ for KdV. Rotation-invariance then follows from the fact that $\sfG(x-y)$ only depends on the difference of its arguments.

\paragraph{Lie-Poisson equations.} The dual space of any Lie algebra admits a Poisson bracket \cite[sec.\ I.4]{Khesin}, and the ensuing equations of motion --- once a Hamiltonian has been chosen --- are known as {\it Lie-Poisson equations} \cite{Marsden}. In the Virasoro case, given the Hamiltonian functional $H$, Lie-Poisson evolution reads $(\dot p,\dot c)=\ad^*_{\cW[p]}(p,c)$ in terms of the velocity \eqref{wmap} and the coadjoint representation \eqref{adot}. One thus has $\dot c=0$ and
\be
\dot p
+\cW[p] p'+2\cW[p]'p-\frac{c}{12}\cW[p]'''
=
0.
\label{lip}
\ee
In what follows we systematically assume that the (constant) central charge does not vanish: $c\neq0$. When $\cW[p]=p$, \eqref{lip} reduces to the KdV equation \cite{Khesin}. In the CH case \cite{Camassa}, the evolution of $p$ is given by a non-local integro-differential equation, while its Lagrangian (Euler-Poincar\'e) counterpart is local (see eqs.\ \eqref{noloch}-\eqref{che} below). Accordingly, for future reference, let us rewrite Lie-Poisson dynamics in terms of velocity instead of momentum: letting $p=\cI[u]$ with $\cI$ the inverse of the velocity map \eqref{wmap}, eq.\ \eqref{lip} is equivalent to
\be
\cI[\dot u]
=
\ad^*_u(\cI[u]),
\qquad
\text{\ie}\qquad
\cI[\dot u]
=
-u(\cI[u])'-2u'\cI[u]+\frac{c}{12}u'''.
\label{udot}
\ee
We will rely on this reformulation in section \ref{sepege} to analyse the CH equation.

\subsection{Exact reconstruction of amenable travelling waves}
\label{secrec}

Having reviewed the basics, let us now prove that the reconstruction of travelling waves is integrable. To show this, we first recall the definition of reconstruction, then introduce a notion of drift velocity, and eventually focus on amenable waves. The derivation parallels \cite[sec.\ 4.1]{Oblak:2020jek}, but is much more general and relies on the velocity map \eqref{wmap}. We end by revisiting reconstruction from the standpoint of dynamical systems. Topological bifurcations and their relation to Virasoro orbits are studied in section \ref{sebif}.

\paragraph{Generalities on reconstruction.} Consider a Lie group $G$ with algebra $\mg$ and dual $\mg^*$. Lie-Poisson equations in $\mg^*$ can be seen, quite generally, as a reduction of dynamics on the cotangent bundle $T^*G$ (see \eg \cite{Marsden,Ratiu} or \cite[app.\ B]{Oblak:2020jek}). Reconstruction goes the other way around, producing the motion $g_t\in G$ that is `sourced' by the momentum $p(t)\in\mg^*$ according to\footnote{The time-dependence of $g_t$ is indicated by a subscript instead of parentheses, to stress the eventual asymmetric roles of $t$ and $x$. In particular, $g_t$ has nothing to do with the time derivative $\der_tg_t\equiv\dot g_t$.}
\be
\dot g_t\,g_t^{-1}
=
\dd H_{p(t)}.
\label{recoco}
\ee
In rigid bodies, for instance, the Lie-Poisson dynamics of angular momentum coincides with the Euler equations of a free-falling top, and eq.\ \eqref{recoco} yields the time-dependent orientation of the body \cite{Montgomery}. Similarly, the local velocity of ideal fluids satisfies Euler's equation $\dot\bu+\bu\cdot\nabla\bu=-\nabla P$ ($P$ being pressure), whose reconstruction describes the deformations of the fluid as a whole, hence also the motion of fluid particles such that $\dot\bx(t)=\bu(\bx(t),t)$ \cite{Arnold}. We now discuss reconstruction in detail for the Virasoro group.

Let a wave profile $p(x,t)$ solve the Lie-Poisson equation \eqref{lip} for some Hamiltonian $H$; at any time $t$, it defines a $2\pi$-periodic function $p(t)(x)\equiv p(x,t)$. Seeing Lie-Poisson dynamics as a reduction of the equations of motion in $T^*\hDiff$, the time evolution of the configuration $g_t\in\Diff$ is given by eq.\ \eqref{recoco}. In the case at hand, this yields\footnote{As Virasoro extends $\Diff$, the full reconstruction condition contains a central piece \cite{HolmTyr,Oblak:2020jek}. What we write in \eqref{rec} is only the {\it centreless} part of reconstruction, which suffices for our purposes.}
\be
\big(\dot g_t\circ g_t^{-1}\big)(x)
=
\frac{\delta H}{\delta p(x)}
\stackrel{\text{\eqref{wmap}}}{=}
\cW[p(t)](x),
\label{rec}
\ee
to which we shall refer as the  {\it reconstruction condition} associated with $p(x,t)$.
Note for future reference that the identity $\der_tg\circ g^{-1}+g'\circ g^{-1}\der_tg^{-1}=0$ allows one to rewrite \eqref{rec} as
\be
\der_t(g_t^{-1})(x)+\cW[p(t)](x)\,\der_xg_t^{-1}(x)=0.
\label{advec}
\ee
This is an advection equation $\der_tg^{-1}+u\,\der_xg^{-1}=0$ with local velocity $u(x,t)\equiv\cW[p(t)](x)$.

From the perspective of hydrodynamics, it is tempting to interpret the reconstruction condition \eqref{rec} as an equation of motion for `fluid particles'. Indeed, any reconstruction $g_t$ determines a path $x(t)=g_t(g_0^{-1}(x_0))$, describing a particle initially at $x_0$ with velocity
\be
\dot x(t)
=
\cW[p(t)](x(t))
\equiv
u\big(x(t),t\big).
\label{xdot}
\ee
When $p(x,t+T)=p(x,t)$ is periodic in time, one may therefore define the {\it drift velocity}
\be
v_{\text{Drift}}
\equiv
\lim_{t\to+\infty}\frac{x(t)-x(0)}{t}.
\label{vidi}
\ee
This can also be written as a quotient $\Delta\phi/T$, where $\Delta\phi$ is the Poincar\'e rotation number of $g_T\circ g_0^{-1}$. One can show that $\Delta\phi$ is in fact a sum of geometric phases \cite{Oblak:2020jek}, but we will not pursue this line of thought here. What matters for our purposes is that, as we shall argue, $v_{\text{Drift}}$ and $\Delta\phi$ are sensitive to the topology of the Virasoro coadjoint orbit of $p$.

Note that, despite striking similarities, eq.\ \eqref{xdot} does {\it not} coincide with the velocity of realistic fluid particles. This mismatch occurs even in shallow water, where, to first approximation, fluid motion really takes place in one (horizontal) direction \cite{Korteweg,Camassa}. The difference lies in a choice of reference frames
and in numerical factors that cannot be absorbed by redefinitions (see \eg \cite[sec.\ 3.3]{Oblak:2020jek}). In particular, the drift velocity \eqref{vidi} does look suspiciously close to Stokes drift \cite{Ursell}, but also differs from it in key details. Accordingly, our point of view will be that reconstruction and eq.\ \eqref{xdot} exhibit rich and subtle geometric phenomena affecting nonlinear waves, but we make no claim as to the relevance of these considerations for actual fluids.\footnote{This mismatch does not occur in 2D or 3D ideal fluids with fixed boundaries, where reconstruction coincides {\it exactly} with the actual fluid motion \cite{Arnold}. We will study such setups elsewhere.} It would be very interesting indeed to devise realistic hydrodynamics experiments where our arguments apply.

\paragraph{Reconstruction of travelling waves.} A {\it travelling wave} is a solution of the Lie-Poisson equation \eqref{lip} whose shape is constant in time, so that $p(x,t)=p(x-vt)$ for some velocity $v$. We call the wave {\it amenable} if it can be mapped on a uniform profile $k$ under the Virasoro coadjoint action \eqref{cog}, \ie if there exists a diffeomorphism $g_0$ such that
\be
p(x)
=
(g_0\cdot k)(x)
\stackrel{\text{\eqref{cog}}}{=}
\big((g_0^{-1})'(x)\big)^2k
-
\frac{c}{12}\bigg[\frac{(g_0^{-1})'''}{(g_0^{-1})'}-\frac{3}{2}\bigg(\frac{(g_0^{-1})''}{(g_0^{-1})'}\bigg)^2\bigg]\Bigg|_x.
\label{gok}
\ee
We will assume that the constant $k$ is {\it generic}, \ie that there is no positive integer $n$ such that $k=-n^2c/24$. This implies that the stabilizer of $k$, \ie the set of diffeomorphisms $h$ such that $h\cdot k=k$, consists of rotations $h(x)=x+\theta$ \cite{Witten:1987ty,Balog:1997zz}.

As we now show, the fact that $p(x,t)$ is an amenable travelling wave simplifies the reconstruction condition \eqref{rec} drastically. The argument is similar to \cite[sec.\ 4.1]{Oblak:2020jek}, but the presence of an arbitrary Hamiltonian (instead of that of KdV) makes the formulas more involved and more general, and exhibits the importance of the velocity map \eqref{wmap}.

Note first that the reconstruction condition \eqref{rec} implies that the solution $p(x,t)$ of the Lie-Poisson equation \eqref{lip} can be written as $p(x,t)=(g_t\cdot k)(x)$,\footnote{Indeed, \eqref{recoco} implies $\der_t(g\cdot k)=\ad^*_{\dot gg^{-1}}(g\cdot k)=\ad^*_{\cW[g\cdot k]}(g\cdot k)$, which is the Lie-Poisson equation \eqref{lip}.} with an initial condition $g_0$ satisfying \eqref{gok}. Since $p$ is an amenable travelling wave, one also has $p(x,t)=p(x-vt)=\big((\cR_{vt}\circ g_0)\cdot k\big)(x)$ in terms of the time-dependent rotation $\cR_{vt}$. Combining the two expressions of $p(x,t)$, one concludes that $g_t^{-1}\circ\cR_{vt}\circ g_0$ stabilizes $k$. As the latter is generic, its stabilizer only contains rotations, so there exists an angle $\theta(t)$ such that
\be
g_t
=
\cR_{vt}\circ g_0\circ\cR_{\theta(t)}.
\label{git}
\ee
Solving the reconstruction condition thus amounts to finding $\theta(t)$, with $\theta(0)=0$. Since eq.\ \eqref{git} is equivalent to $g_t^{-1}(x)=g_0^{-1}(x-vt)-\theta(t)$, the advection equation \eqref{advec} reads
\be
\dot\theta(t)
=
\big[\cW[p(t)](x)-v\big](g_0^{-1})'(x-vt).
\label{ted}
\ee
Now recall that the Hamiltonian is assumed to be rotation-invariant, so $\cW$ satisfies \eqref{wrot}; the fact that $p$ is a travelling wave $p(t)=\cR_{vt}\cdot p(0)$ then allows one to rewrite eq.\ \eqref{ted} as
\be
\dot\theta(t)
=
\big[\cW[p](X)-v\big](g_0^{-1})'(X)
\label{tedd}
\ee
in terms of $X\equiv x-vt$. One can also write this more simply in terms of the local velocity $u=\cW[p]$, namely as $\dot\theta(t)=[u(X)-v](g_0^{-1})'(X)$. In what follows, we will express our conclusions in terms of both momentum $p$ and velocity $u$.

Eq.\ \eqref{tedd} is crucial, for its two sides depend separately on the independent coordinates $t$ and $X$, and must therefore equal a constant $\cV$. (The latter will eventually contribute to the drift velocity \eqref{vidi}, which justifies the notation.) Indeed, eq.\ \eqref{tedd} implies that $u(x)-v$ has no roots and that
\be
g_0^{-1}(x)
=
\phi
+
\int_0^x\frac{\cV\,\dd y}{\cW[p](y)-v}
=
\phi
+
\int_0^x\frac{\cV\,\dd y}{u(y)-v}
\label{gom}
\ee
where $\phi$ is an integration constant. The value of $\cV$ is then fixed by the condition $g_0^{-1}(x+2\pi)=g_0^{-1}(x)+2\pi$, which yields
\be
\boxed{%
\Bigg.
\cV
=
2\pi\left[\int_0^{2\pi}\frac{\dd x}{\cW[p](x)-v}\right]^{-1}
=
2\pi\left[\int_0^{2\pi}\frac{\dd x}{u(x)-v}\right]^{-1}
.}
\label{videf}
\ee
In terms of $\cV$, eq.\ \eqref{tedd} also implies $\theta(t)=\cV t$ owing to the initial condition $\theta(0)=0$. It finally follows that the full reconstructed curve $g_t$, given by eq.\ \eqref{git}, reads
\be
\boxed{%
\Big.
g_t(x)
=
g_0(x+\cV t)+vt.}
\label{exrec}
\ee
This essentially describes the motion of fluid particles satisfying $\dot x(t)=u(x(t),t)$, according to $x(t)=g_t(g_0^{-1}(x_0))=g_0(g_0^{-1}(x_0)+\cV t)+vt$. In particular, the drift velocity defined in eq.\ \eqref{vidi} can be written in closed form:
\be
\boxed{%
\Big.
v_{\text{Drift}}
=
v+\cV
\qquad
\text{(amenable travelling waves)}.}
\label{vivi}
\ee
The function $g_0(x)$ and the constant $\cV$ are entirely fixed by the shape of the wave and its velocity. Once the travelling wave solution $p(x-vt)$ is known, eqs.\ \eqref{gom}-\eqref{exrec} thus provide its {\it exact} Lie-Poisson reconstruction. Equivalently, from a Lagrangian perspective (which will be more convenient for peakons), eqs.\ \eqref{gom}-\eqref{exrec} describe the Euler-Poincar\'e reconstruction of a velocity vector field $u(x-vt)$ solving eq.\ \eqref{udot}.

\paragraph{Reconstruction as a dynamical system.} Our derivation of eqs.\ \eqref{exrec}-\eqref{vivi}, based on amenability, may have obscured the fact that the reconstruction of travelling waves is, at heart, an elementary problem. Indeed, when $u(x,t)=u(x-vt)$ is the velocity field of a travelling wave, eq.\ \eqref{xdot} for `fluid particles' is equivalent to $\dot X=u(X)-v\equiv U(X)$ in terms of $X\equiv x-vt$. This is a simple one-dimensional dynamical system (\ie the flow of a one-dimensional vector field) whose general solution is implicitly given by
\be
\int_{X(0)}^{X(t)}\frac{\dd X}{u(X)-v}
=
t.
\label{implis}
\ee
When $u(X)-v$ has no roots, as was implied above by eq.\ \eqref{tedd}, this solution coincides with that specified by \eqref{exrec}. In particular, upon dividing \eqref{implis} by $t$ and taking the late-time limit, the definition \eqref{vidi} of drift velocity and the $2\pi$-periodicity of $u(x)$ yield
\be
1
=
\lim_{t\to+\infty}\frac{X(t)-X(0)}{t}\,\frac{1}{X(t)-X(0)}\int_{X(0)}^{X(t)}\frac{\dd X}{u(X)-v}
=
\frac{v_{\text{Drift}}-v}{\cV}
\ee
in terms of the constant \eqref{videf}. This manifestly reproduces eq.\ \eqref{vivi} without any mention of Virasoro orbits. Furthermore, the (implicit) solution \eqref{implis} is more general than \eqref{exrec}, since it even holds when $u(X)-v$ has roots (provided $X(0)$ is not a root). One might therefore wonder why we adopted such a convoluted point of view to derive a comparatively elementary fact; the reason, as we now argue, is that our approach exhibits the role of Virasoro geometry in one-dimensional dynamics.

\subsection{Topological bifurcations}
\label{sebif}

Our proof of integrable reconstruction relied on the assumption that $p(x,t)$ is amenable. This is by no means guaranteed: many coadjoint orbits of the Virasoro group have no uniform representative whatsoever \cite{Lazutkin,Kirillov,Witten:1987ty,Balog:1997zz}. Accordingly, we now explore in detail the relation between orbits and reconstruction. This will provide a link between amenability and the (absence of) roots of the function $u(x)-v$, and will motivate the definition of certain bifurcations reflecting sudden changes in orbital topology. Such transitions are in fact saddle-node bifurcations of the flow equation $\dot X=U(X)$, so a corollary of our construction will be a dictionary relating one-dimensional dynamics to Virasoro orbits.

\paragraph{Roots and amenability.} An immediate consequence of section \ref{secrec} is the following: if $p(x-vt)$ is a (generic) amenable travelling wave solution of eq.\ \eqref{lip}, then the function $\cW[p](x)-v=u(x)-v$ has no roots. Indeed, this is guaranteed by eq.\ \eqref{tedd}, which follows from the assumption of amenability and becomes nonsensical when $\cW[p](x)-v$ has roots. The contraposition of this statement is often more useful in practice: if $\cW[p](x)-v$ does have roots {\it and} $p(x-vt)$ is known to solve eq.\ \eqref{lip}, then $p(x)$ cannot be amenable.

We now show that the converse implication also holds --- namely that any travelling wave for which $u(x)-v$ has no roots is necessarily amenable. Indeed, consider a travelling wave $p(x-vt)$ that solves the Lie-Poisson equation \eqref{lip}, and write $U(x)\equiv u(x)-v$:
\be
\label{lipa}
Up'+2U'p-\frac{c}{12}U'''=0.
\ee
Multiplying this by $U$ to recognize a total derivative, one can integrate over $x$ to find $U^2p=A+\frac{c}{12}U''U-\frac{c}{24}U'^2$ for some constant $A$. {\it Provided $U$ is non-vanishing everywhere}, one may divide both sides by $U^2$ and thereby recognize the statement $p=g_0\cdot k$ in terms of the coadjoint action \eqref{gok}, with $g_0$ given by \eqref{gom} and
\be
k\cV^2
=
A
\equiv
\big(u(x)-v\big)\Big[\big(u(x)-v\big)p(x)
-\frac{c}{12}u''(x)\Big]
+\frac{c}{24}u'(x)^2.
\label{kav}
\ee
This proves that any travelling wave $p(x-vt)$ for which $u(x)-v$ has no roots is automatically amenable, with a constant $k$ that can be deduced from Lie-Poisson dynamics through eq.\ \eqref{kav}. In conclusion:
\begin{center}
\begin{framed}
\begin{minipage}{.9\textwidth}
\centering
{\it A travelling wave $p(x-vt)$ solving the Lie-Poisson equation \eqref{lip} is amenable if and only if the function $\cW[p](x)-v=u(x)-v$ has no roots.}
\end{minipage}
\end{framed}
\end{center}
 In the KdV equation, for which $\cW[p]=p$, this result is especially transparent and provides a straightforward orbital analysis of cnoidal waves \cite{Oblak:2019llc,Oblak:2020jek}.

\paragraph{Topological bifurcations and resonances.} Having found that travelling waves are amenable if and only if $u(x)-v$ has no roots, one may wonder what happens when $u(x)-v$ {\it does} have roots. In that case, the solution of the reconstruction condition \eqref{rec} is no longer given by the simple expression \eqref{exrec}, and one must rely instead on the implicit expression \eqref{implis}. The latter states that the motion of $X(t)$ is typically exponential in the vicinity of any root of $u(X)-v$,\footnote{This argument fails for degenerate roots, where $u(x_*)-v=u'(x_*)=0$, occurring at the transition between amenable and non-amenable waves. We briefly return to that case at the very end of this section.} so one has $x(t)\sim vt+x_*+\lambda e^{u'(x_*)t}$ at late times in terms of some stable root $x_*$ (where $u'(x_*)<0$). In particular, $g_t\circ g_0^{-1}$ consists of time-dependent exponential `contractions' towards the stable roots of $u(x-vt)-v$, and the drift velocity \eqref{vidi} coincides with that of the wave:
\be
\boxed{%
\Big.
v_{\text{Drift}}
=
v
\qquad
\text{(non-amenable travelling waves).}}
\label{vivid}
\ee
Equivalently, writing $v_{\text{Drift}}=\Delta\phi/T$ in terms of the period $T=2\pi/|v|$ and the rotation number $\Delta\phi$ of $g_T\circ g_0^{-1}$, one has $\Delta\phi=\pm2\pi$: all fluid particles rotate, on average, by one full circle during one period. This should be contrasted with the amenable drift \eqref{vivi}, where $\cV\neq0$ depends continuously on the wave profile and $\Delta\phi\neq2\pi$ as a result.

The behaviour of drift velocity thus depends delicately on wave amenability. This stems from topology: the stabilizer of (generic) amenable waves is a U(1) group of rotations, while the stabilizer of non-amenable orbits turns out to be a non-compact one-parameter group \cite{Witten:1987ty,Balog:1997zz}. Amenable and non-amenable Virasoro orbits are thus homotopically inequivalent, and this distinction translates into the difference between $v_{\text{Drift}}\neq v$ and $v_{\text{Drift}}=v$. In particular, when the profile $p(x-vt)$ (or the local velocity $u(x-vt)$) is varied continuously, transitions between amenable and non-amenable waves may occur, producing what we call {\it topological bifurcations}.

Similarly to our remark at the end of section \ref{secrec}, this statement can also be phrased in much simpler terms: topological transitions of the equation of motion $\dot X=u(X)-v$ are accompanied by the (dis)appearance of roots, and are thus saddle-node bifurcations analogous to those of the Adler equation \cite{Adler}. From that perspective, eq.\ \eqref{vivid} states that particle motion `resonates' with non-amenable waves. This is quite elementary, so our point is not so much to solve the equation $\dot X=u(X)-v$ as to stress its link with the topology of Virasoro orbits when $u(x-vt)$ satisfies a Lie-Poisson equation. For instance, one may ask under what conditions the one-dimensional dynamical system $\dot X=U(X)$ is equivalent to the reconstruction of a rotation-invariant Lie-Poisson Hamiltonian at central charge $c$. The answer, owing to eq.\ \eqref{lipa}, is that there must exist a velocity $v$ such that $U$ belongs to the stabilizer of $\cI[U+v]$, where $\cI$ is the inverse of the velocity map \eqref{wmap}. When this is the case, the roots of $U(X)$ determine the type of orbit of $\cI[U+v]$: if $U$ has no roots, then $\cI[U+v]$ is amenable; if $U$ has $2n$ simple roots per wavelength, then $\cI[U+v]$ belongs to a hyperbolic orbit with winding $n$ \cite{Balog:1997zz}.

Finally, a note on the Virasoro orbits occurring {\it exactly} at a topological bifurcation point. Bifurcations occur when the function $u(X)-v$ has {\it degenerate} roots, where $u'(X)$ vanishes in addition to $u(X)-v$. The corresponding orbit is typically amenable but non-generic, with an {\it exceptional} uniform representative $k=-n^2c/24$ specified by some positive integer $n$ and stabilized by a three-dimensional group \cite{Witten:1987ty,Balog:1997zz}. Since all other profiles (amenable or not) have smaller, one-dimensional, stabilizers, one can think of bifurcations as points of enhanced symmetry.

\Needspace{10\baselineskip}
\section{Drift and reconstruction of periodic peakons}
\label{sech}

Similarly to KdV \cite{Korteweg}, the Camassa-Holm (CH) equation \cite{Camassa} provides an approximate description of unidirectional waves in shallow water. Here, we study the reconstruction of its simplest periodic travelling waves, namely peakons \cite{Boyd}, whose parameter space is two-dimensional. Using the tools of section \ref{segen}, we will uncover two parameter patches, respectively consisting of amenable and non-amenable profiles, separated by topological bifurcations where drift velocity behaves in a discontinuous manner.

These conclusions are similar to their cnoidal analogues \cite{Oblak:2019llc,Oblak:2020jek}, save for technical simplifications due to the replacement of elliptic functions by hyperbolic functions. They illustrate the results of section \ref{segen} in a system of great interest in physics, so one might hope that they will be useful beyond the context of this work. It would be interesting to find more general applications of our approach, for instance in Lie-Poisson equations whose Hamiltonian is not quadratic, but we will not pursue this here.

This section is organized as follows. We first review the derivation of the CH equation as an Euler-Poincar\'e equation for the Virasoro group, and provide its periodic peakon solutions. Then we find regions of amenability in peakon parameter space, evaluate the resulting drift velocity, and solve reconstruction explicitly thanks to the formulas of section \ref{secrec}. Finally, we compute the uniform representatives of orbits of amenable peakons and relate them to the band structure of a (singular) Kronig-Penney model.

\subsection{Camassa-Holm equation and peakons}
\label{sepege}

Let us start by reviewing the relation between the CH equation and the Virasoro group, and deriving its periodic peakon solutions. Reconstruction is investigated in section \ref{sepek}.

\begin{SCfigure}[2][t]
\centering
\includegraphics[width=0.55\textwidth]{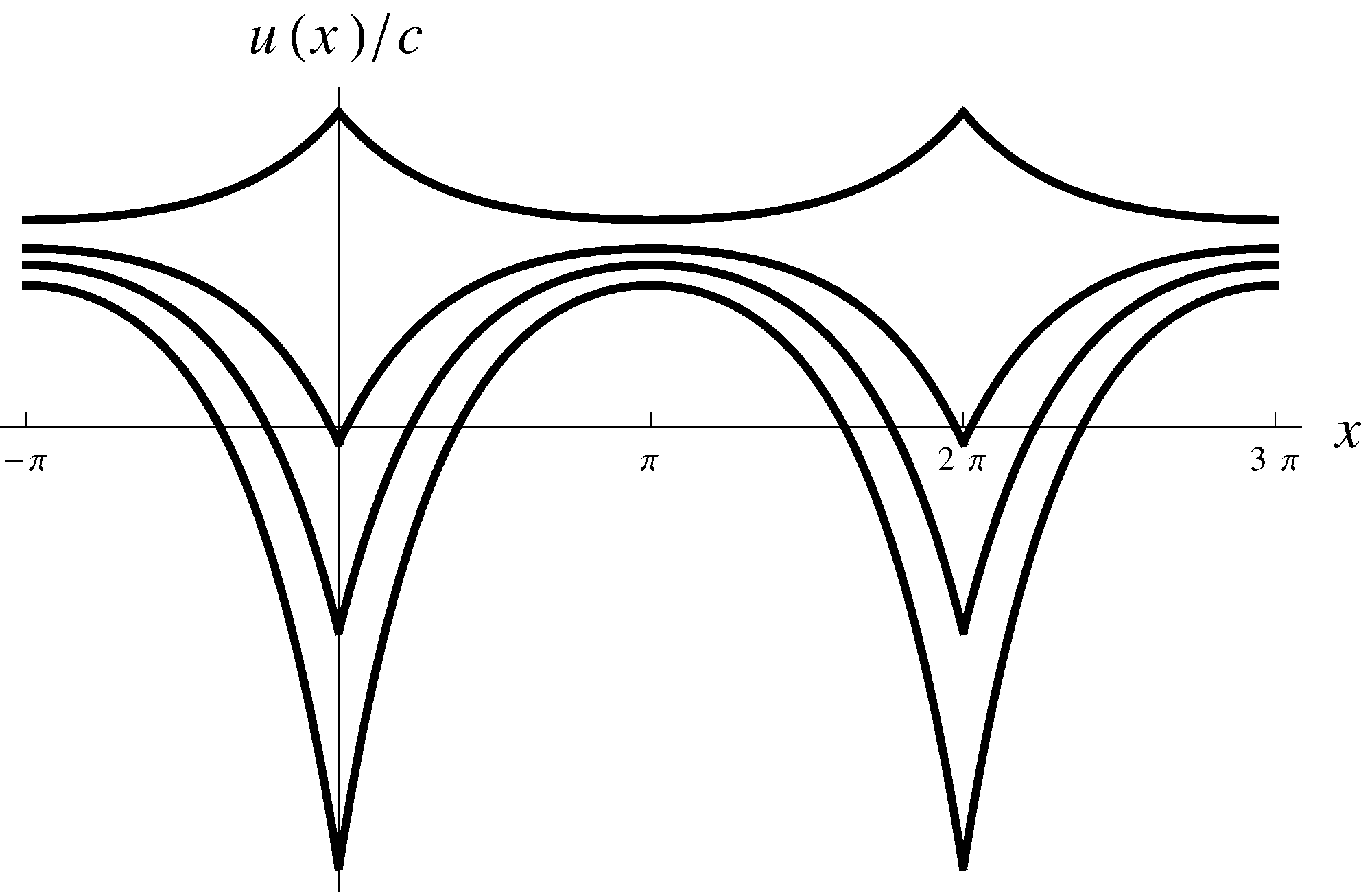}
\caption{Periodic peakons $u(x)/c$ given by \eqref{cosh} with $m=1$ and (top to bottom) $v/c=$ $0.15$, $0.08$, $0.04$, $-0.01$. Up to normalization and additive constants, $u(x)$ coincides with the Green's function \eqref{gif}. Increasing $|v/c-m^2/8|$ increases the amplitude. The profile consists of peaks and troughs when $v/c>m^2/8$, of canyons and hills otherwise.\label{fipik}}
\end{SCfigure}

\paragraph{Inertia, velocity and Hamiltonian.} CH dynamics is a Lie-Poisson equation for the Virasoro group, with an inertia operator $\cI[u](x)=\big(au(x)-bu''(x)\big)\dd x^2$ depending on constants $a,b>0$. This operator is invertible as long as $a>0$,\footnote{The degenerate case $a=0$ yields the Hunter-Saxton equation \cite{Hunter}, which we will not consider.} and one may set $a=1$ without loss of generality upon rescaling time. Having fixed $a\equiv1$, we shall also write $b\equiv1/m^2\equiv\ell^2$ in terms of a dimensionless `mass scale' $m>0$ and its inverse, the `correlation length' $\ell>0$.\footnote{The intuition and terminology for $\ell$ and $m$ stem from quantum field theory, where the correlator of a massive field decays as $e^{-|x-y|/\ell}=e^{-m|x-y|}$. In the present case, the inverse of the inertia operator is a Green's function \eqref{gif} that locally behaves like a decaying exponential of just that form.} In what follows we will initially use both parameters to describe CH dynamics, eventually limiting ourselves to the mass $m$ for convenience. The inverse of $\cI$, \ie the `velocity map' \eqref{wmap}, is a convolution operator
\be
\cW[p](x)
=
\int_0^{2\pi}\dd x\,\sfG(x-y)\,p(y)
\label{vech}
\ee
where $\sfG$ is a $2\pi$-periodic Green's function such that $\sfG(x)-\ell^2\sfG''(x)=\sum_{n\in\ZZ}\delta(x-2\pi n)$ in terms of the delta function on the circle (`Dirac comb'). This yields
\be
\sfG(x)
=
\frac{m}{2}\,
\frac{\cosh\!\big[m(x-\pi)\big]}{\sinh(m\pi)}
\qquad\text{for }x\in[0,2\pi],
\label{gif}
\ee
which is then extended in a $2\pi$-periodic manner over the whole real line. In particular, $\sfG(x)$ is continuous but not differentiable when $x\in2\pi\ZZ$, and reduces to a periodic delta function in the limit $m\to\infty$ ($\ell\to0$). As we shall see, up to normalization and an additive constant, $\sfG(x)$ coincides with the profile of a periodic peakon, so the Green's function of the CH equation is effectively plotted in fig.\ \ref{fipik}.

The Lagrangian kinetic energy associated with the inertia operator $\cI[u]=u-\ell^2u''$ is a local functional $E[u]=\int\dd x(u^2+\ell^2u'^2)/2$. Owing to the velocity \eqref{vech}, the corresponding Hamiltonian functional of $p\equiv\cI[u]$ is non-local:
\be
H[p]
=
\frac{1}{2}
\iint\dd x\,\dd y\,p(x)\sfG(x-y)p(y).
\label{hach}
\ee
(From now on, all integrals over $x$ or $y$ run over $[0,2\pi]$ unless stated otherwise.) This Hamiltonian is rotation-invariant since the Green's function $\sfG(x-y)$ only depends on the difference of its arguments. Furthermore, the velocity map is invertible, so the CH Hamiltonian satisfies both conditions stated in section \ref{sevigo}. In the limit $m\to\infty$ ($\ell\to0$), $H[p]$ becomes local and $\propto\int\dd x\,p(x)^2$, reproducing the standard KdV Hamiltonian. 

\paragraph{Camassa-Holm equation.} In terms of momentum $p$, the Lie-Poisson equation associated with the Hamiltonian \eqref{hach} takes the general form \eqref{lip}. Owing to the velocity operator \eqref{vech}, this is an integro-differential equation:
\be
\dot p(x)
+
\int\dd y\,
\left[%
\sfG(x-y)\,p(y)p'(x)
+2\sfG'(x-y)\,p(y)p(x)
-\frac{c}{12}\sfG'''(x-y)\,p(y)\right]
=
0.
\label{noloch}
\ee
In what follows, we will mostly forget about this formulation of the dynamics and simplify matters by rewriting the system in Lagrangian language. Indeed, in terms of the velocity vector field $u=\cW[p]$, the non-local equation \eqref{noloch} takes the local form \eqref{udot}:
\be
\dot u
+3uu'
-\ell^2\dot u''
-\ell^2uu'''-2\ell^2u'u''-\frac{c}{12}u'''
=
0.
\label{che}
\ee
This is the {\it Camassa-Holm (CH) equation} for $u$ \cite{Camassa,Khesin2003,Khesin}, reducing to KdV when $\ell=0$. As long as $c\neq0$, the actual value of the central charge is irrelevant: one can absorb it by rescaling time and writing \eqref{che} in terms of $u/c$. However, in order to make the relation to the coadjoint representation \eqref{adot} manifest, we will keep the central charge explicit.

Note that the CH equation is not normally presented in the form \eqref{che} in the literature: in order to recover the original expression displayed in \cite{Camassa}, 
\be
\dot U
+2\kappa\frac{\der U}{\der X}
-\frac{\der^2\dot U}{\der X^2}
+3U\frac{\der U}{\der X}
=
2\frac{\der U}{\der X}\frac{\der^2 U}{\der X^2}
+
U\frac{\der^3 U}{\der X^3},
\label{chel}
\ee
one needs to rescale $X\equiv x/\ell$ and define $U\equiv u/\ell+c/(12\ell^3)$, in which case $\kappa=-c/(24\ell^3)$. Note that, in contrast to KdV, the CH equation admits one parameter (namely $\kappa$ in \eqref{chel}, $\ell$ in \eqref{che}, or $m$ in \eqref{gif}) that cannot be absorbed by redefinitions. This parameter will be one of the quantities specifying inequivalent peakons.

In order to keep a manifest relationship between CH dynamics and the Virasoro coadjoint action, we will systematically use the form \eqref{che} of the CH equation (including the explicit central charge). Furthermore, while the equation is normally considered on an entire real line, where it notoriously admits isolated solitons known as {\it peakons} \cite{Camassa}, we will always restrict attention to $2\pi$-periodic local velocities, satisfying $u(x+2\pi)=u(x)$. The corresponding solitons will be {\it periodic} peakons \cite{Boyd}, which we now derive. Their reconstruction, and the ensuing orbital bifurcations, are studied in section \ref{sepek}.

\paragraph{Periodic peakons.} Consider a travelling wave solution $u(x,t)=u(x-vt)$ of the CH equation \eqref{che}. To derive $u$ as a periodic peakon, pick some constant $\lambda$ and let $u(x)=\lambda\cosh[(x-\pi)/\ell]+m^2c/24$ for $x\in[0,2\pi]$, which is then extended to the entire real line in a $2\pi$-periodic way. The function $u(x)$, $x\in\RR$, thus has a discontinuous derivative $u'$ and delta function singularities in $u''$ at all points $x\in2\pi\ZZ$. One readily verifies that the ensuing wave $u(x-vt)$ solves the CH equation \eqref{che} whenever $x-vt\notin2\pi\ZZ$, regardless of $\lambda,v$. However, an extra constraint stems from the requirement that eq.\ \eqref{che} be satisfied, in the sense of distributions, even when $x-vt\in2\pi\ZZ$. This fixes $\lambda=v-m^2c/8$, hence
\be
u(x)
=
\left(v-\frac{m^2c}{8}\right)
\frac{\cosh\!\big[m(x-\pi)\big]}{\cosh(m\pi)}
+
\frac{m^2c}{24}
\qquad\text{for }x\in[0,2\pi],
\label{cosh}
\ee
which is, again, extended to the whole real line in a $2\pi$-periodic manner. We shall refer to the resulting travelling wave $u(x-vt)$ as a {\it (periodic) peakon} or {\it coshoidal wave} \cite{Boyd}. According to eq.\ \eqref{cosh}, peakons are specified by three parameters $(m,v,c)$, but one should keep in mind that the actual value of $c\neq0$ is irrelevant, since the evolution of $u(x,t)/c$ is independent of $c$. Thus, we will consider from now on that periodic peakons are labelled by {\it two} parameters, namely the mass $m$ and the rescaled velocity $v/c$. When $8v/c>m^2$, the profile $u(x)/c$ consists of periodic peaks separated by troughs of length $2\pi$; when $8v/c<m^2$, the peaks point downward and the troughs become hills. The larger the value of $|v-m^2c/8|$, the higher the amplitude of $u$. See fig.\ \ref{fipik}.

Note the similarity between the coshoidal profile \eqref{cosh} and the Green's function \eqref{gif}. This is no accident: the momentum $p$ that corresponds to the local velocity \eqref{cosh} is
\be
p(x)
=
\cI[u](x)
=
u(x)-\frac{u''(x)}{m^2}
=
\frac{m^2c}{24}
+
\left(\frac{2v}{m}-\frac{mc}{4}\right)
\tanh(m\pi)
\sum_{n\in\ZZ}
\delta(x-2\pi n),
\label{sip}
\ee
where the delta functions stem from the discontinuous derivatives of $u(x)$ when $x\in2\pi\ZZ$. Because $p(x)\sim(...)+(...)\delta(x)$, it is clear that $u=\cW[p]=\int\dd y\,\sfG(x-y)p(y)$ will have the form $(...)+(...)\cosh[m(x-\pi)]$, exactly as in eq.\ \eqref{cosh}. In particular, peakon momenta do {\it not} belong to the smooth dual of the Virasoro algebra, and therefore span coadjoint orbits whose elements are singular. Conversely, if one is to map the singular profile \eqref{sip} on some constant $k$, as required for amenable waves, one necessarily has to use diffeomorphisms of the circle that are {\it not} smooth. Reconstruction will make this explicit: when they exist, the diffeomorphisms $g_0(x)$ that turn uniform profiles into peakons are continuous and differentiable, but have discontinuous second derivatives for $x\in2\pi\ZZ$.

\subsection{Peakon reconstruction and topological bifurcations}
\label{sepek}

We now apply the reconstruction tools of section \ref{segen} to periodic peakons and derive the drift velocity \eqref{vivi}. This will result in various expressions that depend parametrically on $(m,v)$. As we shall see, peakons are amenable as long as $v/c$ is large enough (at fixed $m$), but non-amenable when $v/c$ is lower than some ($m$-dependent) critical value. Topological bifurcations separating these two regimes produce a discontinuity in the parameter-dependence of drift velocity. This provides a tractable, yet non-trivial, consistency check of section \ref{segen}, and parallels the analysis of cnoidal waves in \cite{Oblak:2019llc,Oblak:2020jek}.

\paragraph{Amenable peakons and bifurcations.} As shown in section \ref{sebif}, a travelling wave $p(x-vt)$ is amenable if and only if the corresponding local velocity satisfies $u(x)\neq v$ for all $x$. We initially proved this statement for smooth profiles and smooth diffeomorphisms, but the argument extends to distributions such as \eqref{sip} and $C^1$ diffeomorphisms. Now, one readily verifies from eq.\ \eqref{cosh} that $u(x)-v$ has {\it no roots} if and only if
\be
\frac{v}{c}
>
\frac{m^2}{24}
\,
\frac{\cosh(m\pi)-3}{\cosh(m\pi)-1}
\qquad
\text{(amenability condition; $k$ exists)}.
\label{amen}
\ee
Accordingly, a peakon with parameters $(m,v)$ is amenable (Virasoro-equivalent to a generic constant $k$) if its velocity satisfies \eqref{amen}. Conversely, peakons that violate inequality \eqref{amen} are guaranteed to be non-amenable; they belong to hyperbolic Virasoro orbits with unit winding \cite{Balog:1997zz}, just as non-amenable cnoidal waves \cite{Oblak:2019llc}. Peakons for which the inequality \eqref{amen} is replaced by an {\it equality} belong to the exceptional orbit of $k=-c/24$; they are amenable, but not generic.

\begin{SCfigure}[2][t]
\centering
\includegraphics[width=0.55\textwidth]{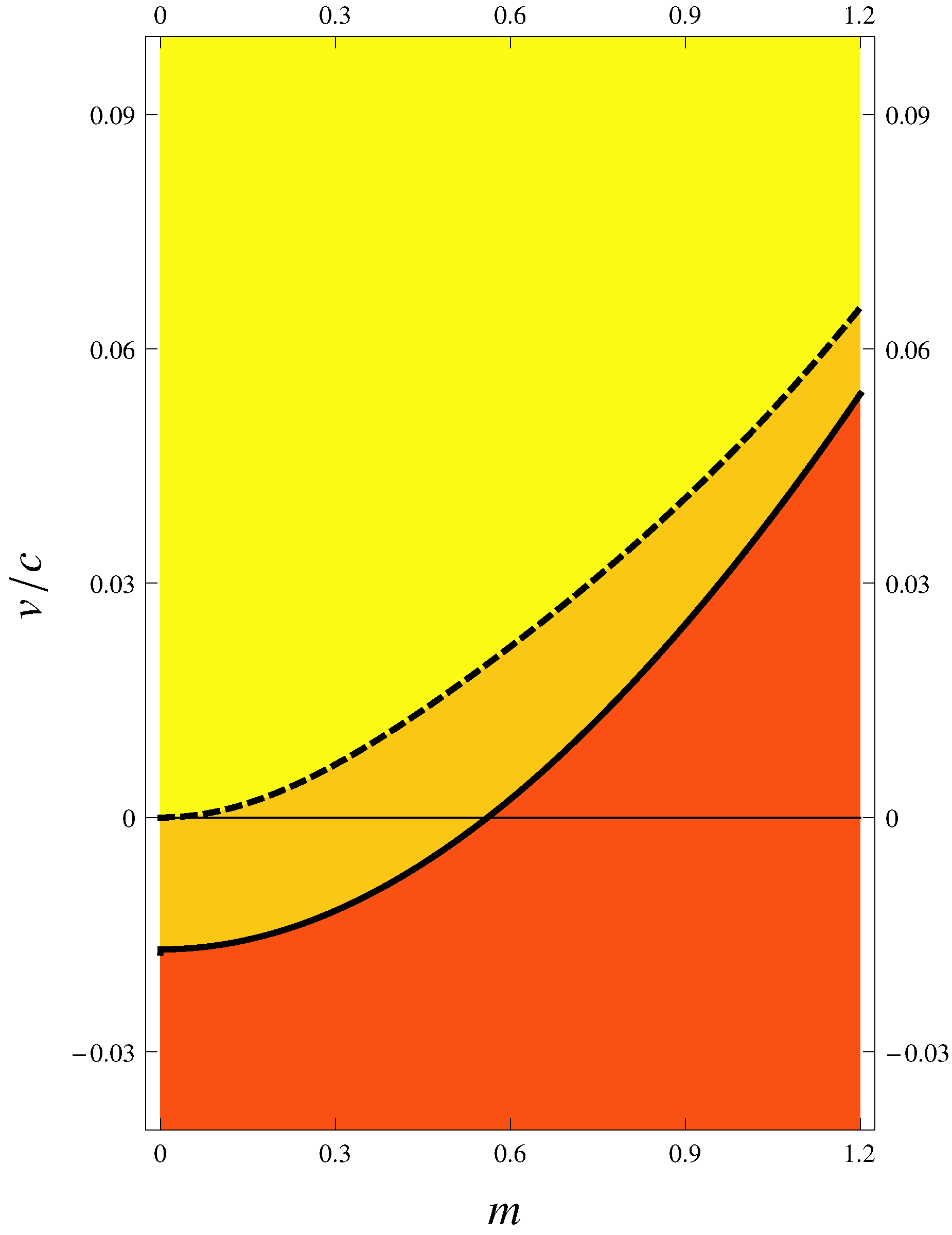}
\caption{A portion of peakon parameter space $(m,v/c)$, with $v=$ velocity, $m=$ mass scale. CH dynamics reduces to KdV (Hunter-Saxton) when $m\to\infty$ ($m\to0$). The {\it red} region, where \eqref{amen} is violated, consists of {\it non-amenable} waves; amenable profiles span the lighter (yellow/orange) regions, where \eqref{amen} holds. Topological bifurcations occur on the line between red and orange patches. The {\it orange} region consists of {\it elliptic amenable} profiles ($-1<24k/c<0$), while the {\it yellow} patch consists of {\it hyperbolic amenable} profiles ($k/c>0$); they are separated by a dashed line ($k=0$). See fig.\ \ref{fikapik} for details on the value of $k$ in the amenable region.\label{fitobif}}
\end{SCfigure}

We have thus identified the line in $(m,v)$ parameter space where peakons undergo topological bifurcations: at fixed $m$, when $v/c$ decreases and crosses the critical value on the right-hand side of \eqref{amen}, Virasoro orbits change from amenable to non-amenable (\ie from orbits that do possess a uniform representative to orbits that lack one). This is plotted in fig.\ \ref{fitobif}, and will be confirmed by the discontinuous behaviour of drift velocity once we turn to the details of reconstruction. In fact, we will see in section \ref{seck} that the picture is somewhat richer, in that even amenable orbits fall in two distinct classes: those with sufficiently low $v/c$ belong to elliptic Virasoro orbits (for which $-1<24k/c<0$), while profiles with high $v/c$ belong to hyperbolic orbits (for which $k/c>0$) \cite{Balog:1997zz}. The transition between those two cases turns out to described by the inequality
\be
\frac{v}{c}
>
\frac{m^2}{24}
\,
\frac{\cosh(m\pi)+3}{\cosh(m\pi)+1}
\qquad
\text{(hyperbolicity condition; $k/c>0$)}.
\label{hyp}
\ee
Finally, it is worth noting that the roots of the velocity function $u(x)$ {\it alone} (as opposed to $u(x)-v$) also contain qualitative information about reconstruction. Indeed, eq.\ \eqref{xdot} implies that roots of $u(x)$ produce oscillations of the position $x(t)$ (since one has both $\dot x>0$ and $\dot x<0$), while the absence of roots of $u(x)$ yields a monotonous behaviour (with constant $\text{sign}(\dot x)$). One readily verifies that the condition for the existence of roots is
$m^2[3-\cosh(m\pi)]/24<v/c<m^2/12$; roughly, oscillations occur at high $m$. Since this is unrelated to the orbit structure of peakons, we will not dwell on it any further.

\begin{SCfigure}[2][t]
\centering
\includegraphics[width=0.55\textwidth]{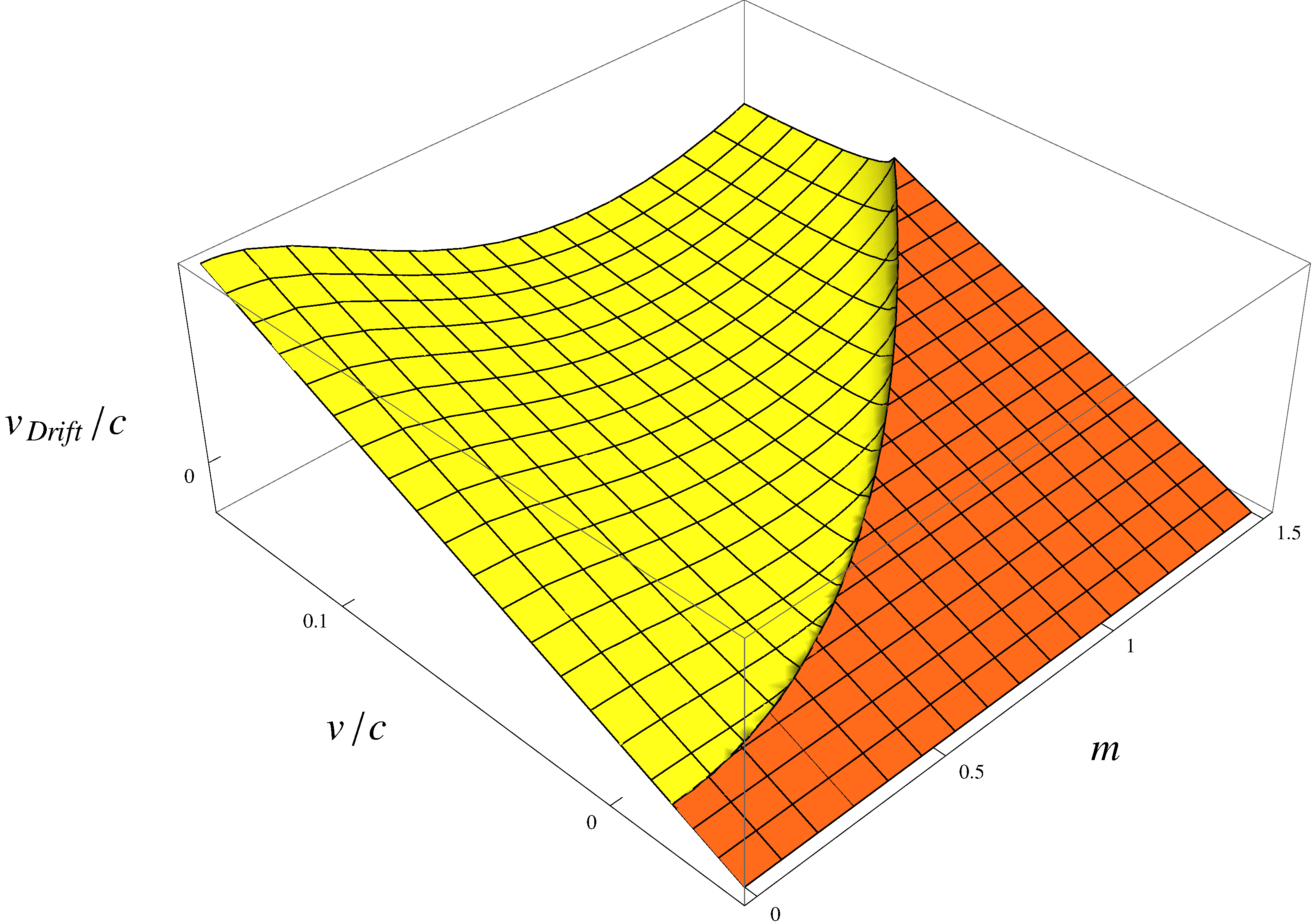}
\caption{The drift velocity \eqref{vidi} of periodic peakons, with parameters $(m,v/c)$. Colours as in fig.\ \ref{fitobif}: amenable profiles span the yellow region, where $v_{\text{Drift}}=v+\cV$ is given by \eqref{vipik}. Non-amenable profiles span the red region, where $v_{\text{Drift}}=v$. Topological bifurcations are manifest. Note that $v_{\text{Drift}}$ is smooth along the border between elliptic and hyperbolic amenable orbits, so (in contrast to fig.\ \ref{fitobif}) this mild transition is not stressed by any change of colour.\label{fidif}}
\end{SCfigure}

\paragraph{Drift velocity.} Sections \ref{secrec}-\ref{sebif} provide the tools needed to evaluate the drift velocity \eqref{vidi} throughout peakon parameter space. First, as shown in eq.\ \eqref{vivid}, particle motion is locked to non-amenable peakons, so $v_{\text{Drift}}=v$ when $(m,v)$ violate the inequality \eqref{amen}. By contrast, when a peakon is amenable, drift velocity is given by eq.\ \eqref{vivi} with $\cV$ defined in \eqref{videf}. This leads to the following expression in the case of peakons:
\be
\frac{2\pi}{\cV}
\stackrel{\text{\eqref{cosh}}}{=}
\frac{2\cosh(m\pi)}{mv-m^3c/8}
\int_0^{m\pi}
\frac{\dd s}{\cosh s-\frac{v-m^2c/24}{v-m^2c/8}\cosh(m\pi)}.
\label{intec}
\ee
This integral is well-defined whenever the amenability condition \eqref{amen} is satisfied. However, its actual evaluation depends delicately on the extra inequality \eqref{hyp}: when \eqref{hyp} holds, the integrand takes the form $(\cosh s+\text{cst})^{-1}$ with $\text{cst}>1$, leading to an inverse hyperbolic function; in the opposite case, one has $-1<\text{cst}<1$, leading to an inverse trigonometric function. This distinction ultimately reflects the difference between hyperbolic and elliptic orbits, to which we shall return in section \ref{seck}. In any case, one can write the result as
\be
v_{\text{Drift}}-v
\stackrel{\text{\eqref{vivi}}}{=}
\cV
=
-\frac{c}{2}\,
\frac{m\pi}{\cosh(m\pi)}\,
\frac{\ds\left[\left(\frac{v}{c}-\frac{m^2}{24}\right)^2\cosh^2(m\pi)-\left(\frac{v}{c}-\frac{m^2}{8}\right)^2\right]^{1/2}}
{\text{artanh}\left[\left(\frac{\left(\frac{v}{c}-\frac{m^2}{24}\right)\cosh(m\pi)+\frac{v}{c}-\frac{m^2}{8}}{\left(\frac{v}{c}-\frac{m^2}{24}\right)\cosh(m\pi)-\frac{v}{c}+\frac{m^2}{8}}\right)^{1/2}\tanh(m\pi/2)\right]}
\label{vipik}
\ee
which strictly holds only for hyperbolic amenable profiles, satisfying inequality \eqref{hyp}, but can be extended to elliptic amenable values thanks to the (branch-independent) property
\be
\frac{\text{artanh}(\sqrt{-|\alpha|}\,\tau)}{\sqrt{-|\alpha|}}
=
\frac{\arctan(\sqrt{|\alpha|}\,\tau)}{\sqrt{|\alpha|}}
\qquad\forall\,\alpha,\tau\in\RR.
\label{taden}
\ee
We have thus found the complete parameter-dependence of drift velocity for solutions of eq.\ \eqref{xdot} in which $u(x-vt)$ is a peakon \eqref{cosh}. The result is plotted in fig.\ \ref{fidif} and behaves smoothly with respect to peakon parameters $(m,v)$, except on the line of topological bifurcations where the derivatives of $v_{\text{Drift}}$ diverge. In particular, nothing dramatic happens on the border \eqref{hyp} between elliptic and hyperbolic amenable peakons; this is no surprise, since all orbits that admit a generic uniform representative are topologically equivalent. Also note that $v_{\text{Drift}}=v$ in the Hunter-Saxton limit $m\to0$, and that, at fixed $m$,
\be
v_{\text{Drift}}
\sim
v
-
\frac{m\pi\tanh(m\pi)}{\log\left[\frac{24}{m^2}\,\frac{\sinh^2(m\pi)}{\cosh(m\pi)}\,\frac{v}{c}\right]}\,v
\qquad
\text{as }v/c\to+\infty.
\label{vidas}
\ee
(The logarithm in the denominator implies $v_{\text{Drift}}\sim v$ at large $v$, but the convergence is exceedingly slow, so eq.\ \eqref{vidas} is useful in comparisons with numerical plots of $v_{\text{Drift}}$.)

\paragraph{Exact reconstruction.} The evaluation of drift velocity is a prerequisite for the exact solution of the equation of motion \eqref{xdot}, since the reconstruction \eqref{exrec} of any amenable travelling wave involves the diffeomorphism \eqref{gom} that maps the wave on a uniform profile. In the case at hand, this yields an integral very similar to \eqref{intec}, resulting in
\be
g_0^{-1}(x)
=
\pi
\frac{\text{artanh}\left[\left(\frac{\left(\frac{v}{c}-\frac{m^2}{24}\right)\cosh(m\pi)+\frac{v}{c}-\frac{m^2}{8}}{\left(\frac{v}{c}-\frac{m^2}{24}\right)\cosh(m\pi)-\frac{v}{c}+\frac{m^2}{8}}\right)^{\!\!1/2}\tanh\big[m(x-\pi)/2\big]\right]}{\text{artanh}\left[\left(\frac{\left(\frac{v}{c}-\frac{m^2}{24}\right)\cosh(m\pi)+\frac{v}{c}-\frac{m^2}{8}}{\left(\frac{v}{c}-\frac{m^2}{24}\right)\cosh(m\pi)-\frac{v}{c}+\frac{m^2}{8}}\right)^{\!\!1/2}\tanh(m\pi/2)\right]}
\qquad\text{for }x\in[0,2\pi]
\label{gopik}
\ee
where we chose the integration constant $\phi=\pi$ in eq.\ \eqref{gom}. This is then extended to the whole real line by defining $g_0^{-1}(x+2\pi)\equiv g_0^{-1}(x)+2\pi$, and is understood to hold throughout the amenable region \eqref{amen} thanks to the identity \eqref{taden}. Note that $g_0^{-1}$ (and its inverse $g_0$) has discontinuities in its second derivative whenever $x\in2\pi\ZZ$; this is a remnant of the discontinuous first derivative of the peakon profile \eqref{cosh}. Thus, amenable peakons can be mapped on a uniform representative $k$ only at the price of using diffeomorphisms that are of class $C^1$, but not $C^2$ (and in particular not smooth).

\begin{figure}[b]
\centering
\includegraphics[width=0.45\textwidth]{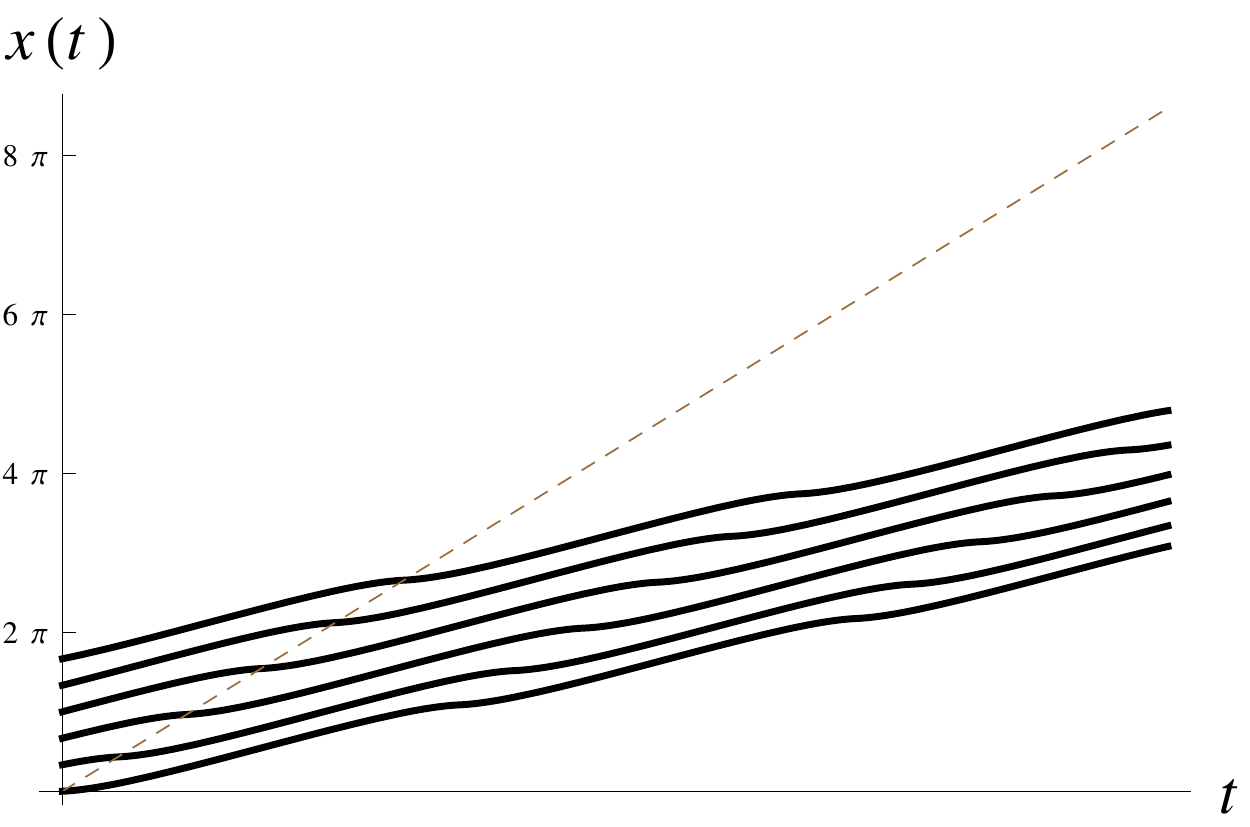}
~~~
\includegraphics[width=0.45\textwidth]{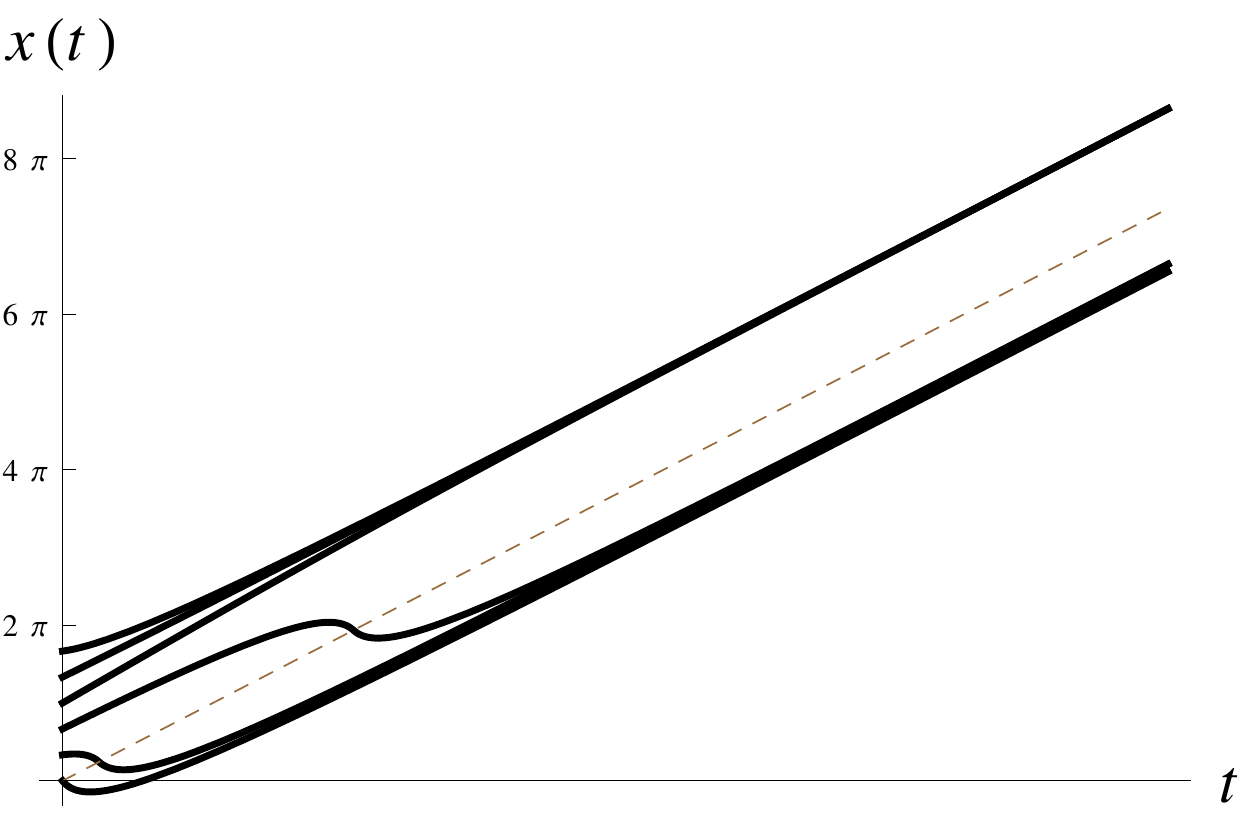}
\caption{Solutions of the equation of motion \eqref{xdot}, $u(x,t)$ being a periodic peakon \eqref{cosh} at $c=m=1$, $v/c=0.09$ (left) and $v/c=0.03$ (right). The former is amenable, the latter is not. In both plots, the dashed line is $x(t)=vt$, confirming that $v_{\text{Drift}}=v$ only for {\it non-amenable} peakons, as in the bifurcation diagram of fig.\ \ref{fidif}.\label{fisols}}
\end{figure}

Using eq.\ \eqref{exrec}, the diffeomorphism \eqref{gopik} yields the explicit motion $x(t)=g_0(g_0^{-1}(x_0)+\cV t)+vt$ of fluid particles whose velocity satisfies eq.\ \eqref{xdot}. This is plotted in fig.\ \ref{fisols} for several choices of initial conditions, along with (numerical) solutions of the equations of motion produced by non-amenable peakons. We stress once more that this can all be deduced from the general solution \eqref{implis} of the flow equation for $u(x)-v$, and is thus trivial from the point of view of one-dimensional dynamical systems. In particular, the discontinuous parameter-dependence of drift velocity in fig.\ \ref{fidif} reflects the usual phase-locking that accompanies saddle-point bifurcations \cite{Adler}. Our main point, which {\it cannot} be deduced from elementary considerations on dynamical systems, is that such bifurcations are closely related to the topology of Virasoro orbits of peakons.

\subsection{Uniformization of peakons}
\label{seck}

To conclude this work, we now compute the constant $k(m,v)$ that is mapped on a peakon momentum \eqref{sip} by the inverse of the diffeomorphism \eqref{gopik}. In particular, we wish to understand in what sense topological (saddle-node) bifurcations between non-amenable and amenable peakons (given by \eqref{amen}) are `more dramatic' than the mild transitions between amenable profiles of elliptic and hyperbolic types (given by \eqref{hyp}). We will see that $k(m,v)$ behaves smoothly throughout the amenability region \eqref{amen} except at its boundary, where it suddenly drops to the exceptional value $-c/24$. This conclusion will be reached in two different ways: first, by evaluating $k$ thanks to eq.\ \eqref{kav}, which follows in this case from CH dynamics and involves the velocity \eqref{vipik}; second, by finding the monodromies of Hill's equations with peakon momenta \eqref{sip} \cite{Lazutkin,Balog:1997zz}. This is related to the band structure of the Kronig-Penney model \cite{Ashcroft}, where transitions between elliptic and hyperbolic profiles signal band edges. Similar considerations on cnoidal waves have been presented in \cite{Oblak:2019llc}.

\begin{SCfigure}[2][t]
\centering
\includegraphics[width=0.55\textwidth]{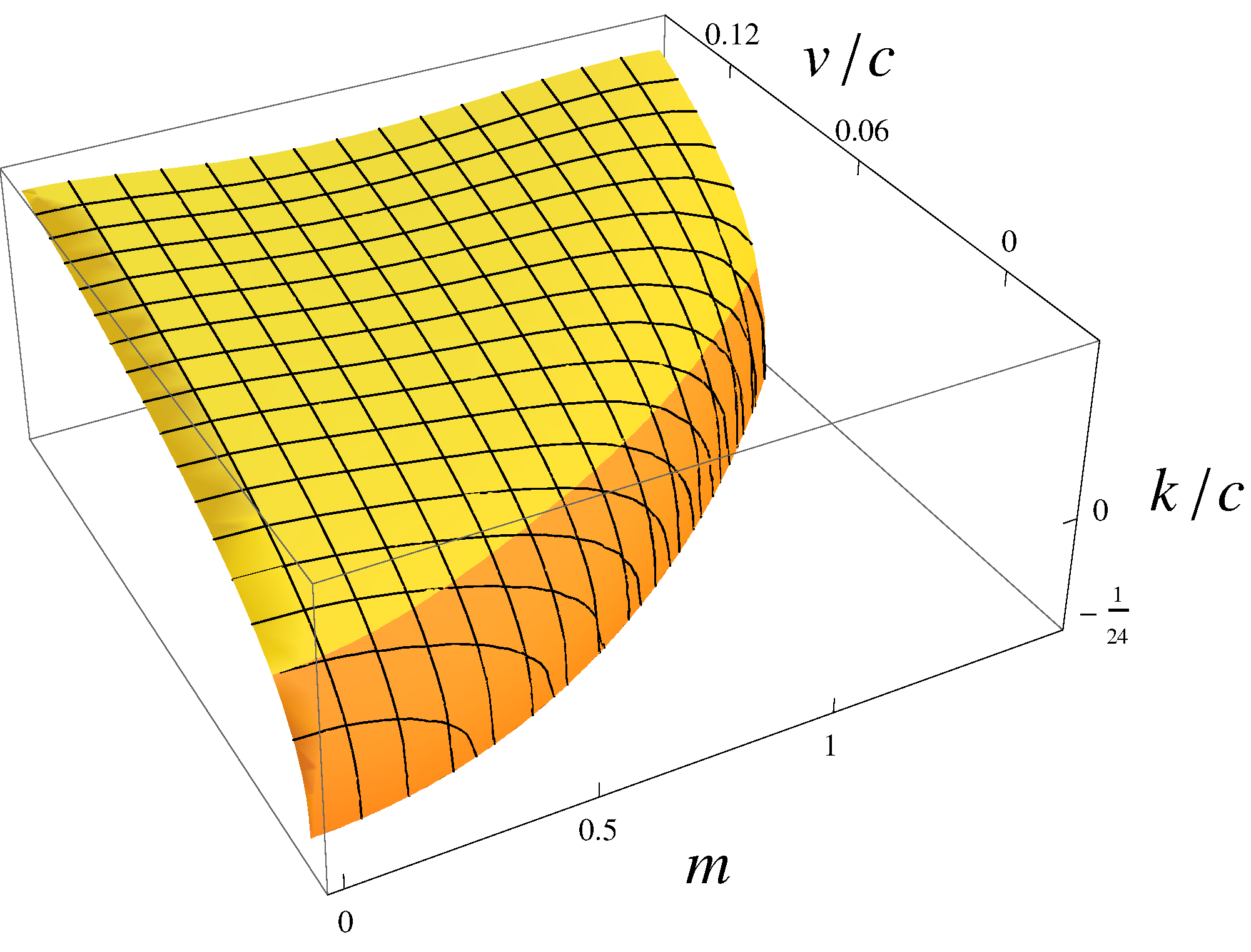}
\caption{The uniform representative $k$ of amenable peakons, given by eq.\ \eqref{kapik} in terms of $(m,v)$. The value $k=0$ occurs on the border between elliptic (orange) and hyperbolic (yellow) amenable peakons, where \eqref{hyp} is saturated. The minimum value of $k/c$ is $-1/24$, occurring precisely on the line of topological bifurcations where the inequality \eqref{amen} is saturated. On that line, derivatives of $k$ with respect to $(m,v)$ diverge.\label{fikapik}}
\end{SCfigure}

\paragraph{Uniformization from CH dynamics.} Eq.\ \eqref{kav} stems from the Lie-Poisson equation \eqref{lip} and yields the uniform value $k$ that corresponds to any amenable travelling wave. Its right-hand side is constant and may therefore be evaluated at any $x$. For peakons, this allows us to take $x\notin2\pi\ZZ$, where the velocity \eqref{cosh} is smooth and the momentum \eqref{sip} reduces to $p(x)=m^2c/24$. Using formula \eqref{vipik} for $\cV$, eq.\ \eqref{kav} thus gives
\be
k
=
\frac{c}{6\pi^2}
\left(\text{artanh}\left[\left(\frac{\left(\frac{v}{c}-\frac{m^2}{24}\right)\cosh(m\pi)+\frac{v}{c}-\frac{m^2}{8}}{\left(\frac{v}{c}-\frac{m^2}{24}\right)\cosh(m\pi)-\frac{v}{c}+\frac{m^2}{8}}\right)^{\!\!\!1/2}\tanh(m\pi/2)\right]\right)^{\!\!\!2}.
\label{kapik}
\ee
Similarly to eq.\ \eqref{vipik}, this expression is strictly valid only for hyperbolic peakons satisfying \eqref{hyp}. Uniform representatives of elliptic peakons (whose parameters satisfy \eqref{amen} but violate \eqref{hyp}) are given by \eqref{kapik} with the replacement $[\text{artanh}(\pm ix)]^2=-[\arctan x]^2$. Note in particular that $k/c>0$ for hyperbolic amenable profiles, while $k/c<0$ for elliptic ones. The transition between those two cases, at $k=0$, occurs on the line where \eqref{hyp} is saturated. Also note that the minimum value of $k/c$ is $-1/24$, occurring precisely at the topological bifurcation where \eqref{amen} is saturated. This is depicted in fig.\ \ref{fikapik} and confirms the relation between coadjoint orbits and saddle-node bifurcations of reconstructed dynamics.

Eq.\ \eqref{kapik} is similar to its analogue for cnoidal waves \cite[eq.\ (27)]{Oblak:2019llc}. In both cases, wave profiles may belong to any hyperbolic amenable orbit ($k/c>0$) or any elliptic orbit with unit winding ($-1/24<k/c<0$), with a smooth transition in between. Furthermore, topological bifurcations occur exactly when $k=-c/24$, \ie at the first `exceptional value' of $k$, where the parameter-gradient of $k$ diverges. Non-amenable profiles span, in both cases, the $n=1$ hyperbolic orbit of the Virasoro group (which indeed has no uniform representative), since $u(x)-v$ has precisely two simple roots per wavelength in the non-amenable region. A notable difference between peakons and cnoidal waves is that the former always have $k/c\geq-1/24$, while cnoidal waves can have arbitrary $k$ \cite{Oblak:2019llc}.

\paragraph{Uniformization from Hill's equation.} For completeness, we now repeat the computation of the constant \eqref{kapik} using a different but more standard method. Namely, given the momentum $p(x)=p(x+2\pi)$, we think of it as a periodic `potential' for {\it Hill's equation} \cite{Balog:1997zz}
\be
-\frac{c}{6}\psi''(x)
+p(x)\psi(x)
=
0,
\label{hill}
\ee
where $\psi(x)$ is a real function on $\RR$. One may roughly see \eqref{hill} as a Schr\"odinger equation, except that $\psi$ is not complex and generally neither periodic, nor square-integrable on $\RR$. Given two linearly independent solutions of \eqref{hill}, say $\psi_1$ and $\psi_2$, the periodicity of $p(x)$ implies the existence of a monodromy matrix $\sfM\in\text{SL}(2,\RR)$ such that
\be
\begin{pmatrix}
\psi_1(x+2\pi) \\ \psi_2(x+2\pi)
\end{pmatrix}
=
\sfM\cdot
\begin{pmatrix}
\psi_1(x) \\ \psi_2(x)
\end{pmatrix}
\qquad\forall\,x\in\RR.
\label{monod}
\ee
It turns out that the conjugacy class of $\sfM$ in $\text{SL}(2,\RR)$ contains nearly all the information needed to deduce the coadjoint orbit of $p(x)$. In particular, when $p(x)$ is amenable, the corresponding uniform profile $k$ follows from the trace \cite[eq.\ (7.22)]{Oblak:2016eij}
\be
\text{Tr}(\sfM)
=
2\cosh\Big(2\pi\sqrt{6k/c}\Big).
\label{mk}
\ee
This formula holds for both positive and negative $k/c$. In the former case, $\text{Tr}(\sfM)\geq2$ and $\sfM$ is {\it hyperbolic}; conversely, when $k/c<0$, one has $-2<\text{Tr}(\sfM)<2$ and $\sfM$ is {\it elliptic}. Note that, as a result, a necessary condition for amenability is $\text{Tr}(\sfM)\geq-2$.

We will not review the relation between Hill's equation and Virasoro orbits in any depth here, referring to \cite[sec.\ II.2]{Khesin}, \cite{Guieu} or \cite[chaps.\ 6-7]{Oblak:2016eij} for details. Instead, let us bluntly apply the method to periodic peakons, whose momenta are given by \eqref{sip}. Consider therefore the family of Hill's equations of the form
\be
-\psi''(x)
+\Big(A+B\sum_{n\in\ZZ}\delta(x-2\pi n)\Big)\psi(x)
=
0
\label{geh}
\ee
where $A,B$ are real parameters which, for peakons, read $A=m^2/4$ and $B=\big[12v/(mc)-3m/2\big]\tanh(m\pi)$. This equation is readily solved by elementary arguments from one-dimensional quantum mechanics: in each open interval $(2\pi n,2\pi(n+1))$, the linearly independent solutions are exponentials $e^{\pm\sqrt{A}\,x}$, while the delta functions at $x\in2\pi\ZZ$ enforce relations between the first derivatives of $\psi$ on the left and on the right of each point $x=2\pi n$. Requiring in addition that $\psi$ be continuous, one finds $\text{Tr}(\sfM)=2\cosh(2\pi\sqrt{A})+(B/\sqrt{A})\sinh(2\pi\sqrt{A})$, readily providing the value of $k$ through eq.\ \eqref{mk}. Since $A=m^2/4$ and $B=\big[12v/(mc)-3m/2\big]\tanh(m\pi)$, this expresses $k$ as a function of peakon parameters $(m,v)$, and a few elementary algebraic manipulations reproduce the earlier result \eqref{kapik}, confirming its validity. This exhibits, once more, that saddle-node bifurcations of reconstruction are topological transitions of Virasoro orbits.

The monodromy method based on the Schr\"o\-din\-ger-\-like equation \eqref{hill} is manifestly related to band structures of one-dimensional periodic potentials. Indeed, one can think of $\psi$ as a wavefunction whose energy is hidden in the parameter $A$ of \eqref{geh}, and the potential consisting of periodic delta functions identifies the system with (a limit of) the Kronig-Penney model \cite{Ashcroft}. From that perspective, requiring the monodromy to be elliptic amounts to the condition that $\psi$ be a Bloch state with well-defined crystal momentum $q=\sqrt{-6k/c}$. It follows that both topological bifurcations (where $k$ becomes complex) and the mild transition between elliptic and hyperbolic profiles (where $k$ changes sign) signal band edges --- and both are equally dramatic. What makes the constant $k$ smooth at one of the two edges, as opposed to the singularities of $q$ at both edges, is the fact that $k\propto q^2$ is real even when $q$ becomes purely imaginary. The same observations apply to cnoidal waves and the related Lam\'e band structure \cite{Oblak:2019llc}.

\section*{Acknowledgements}

I am grateful to B.\ Estienne and G.\ Kozyreff for numerous discussions and collaboration on related subjects. This work is supported by the ANR grant {\it TopO} No.\ ANR-17-CE30-0013-01, and by the European Union’s Horizon 2020 research and innovation programme under the Marie Sk\l{}odowska-Curie grant agreement No.\ 846244.

\Needspace{10\baselineskip}
\addcontentsline{toc}{section}{References}

\end{document}